\documentclass[preprint]{imsart}

\usepackage{amsthm,amsmath,graphicx,float,longtable,indentfirst,textcomp}

\RequirePackage[OT1]{fontenc}
\RequirePackage[numbers]{natbib}
\RequirePackage[colorlinks,citecolor=blue,urlcolor=blue]{hyperref}

\startlocaldefs
\endlocaldefs

\begin{document}

\begin{frontmatter}

\title{One More Way to Encrypt a Message}
\runtitle{ One More Way to Encrypt a Message}

\author{\fnms{Irina} \snm{Pashchenko}\ead[label=e1]{pashchenko.irina.16@gmail.com}}
\runauthor{Irina Pashchenko }
\address{\printead{e1}}

\begin{abstract}
This work describes an example of an application of a novel method for symmetric cryptography. Its purpose is to show how a regular message can be encrypted and then decrypted in an easy, yet secure way. The encrypting method introduced in this work is different from others because it involves decimals as well as integers, encrypting the same initial message differently every time, and inserting misleading digits into every encrypted message, thus making the task of breaking the code even harder. A C++ program was written to support each chapter.
\end{abstract}


\tableofcontents

\end{frontmatter}

\section{Introduction}

Cryptography is a science that deals with the scheme and application of secrecy systems  \cite{ros}. 

Encrypting is a process of transforming a message into such a form that the message cannot be read in a normal way. The procedure is done for the purpose of hiding a message from anybody who is not intended to read it. A method which is chosen for the purpose of encryption is called a cipher \cite{ros}.

Normally, a secret code is used to encrypt a message. A person, who is authorized to read the message later, should be provided with the code in a safe place ahead of time. Then, an original message is created and encrypted with the above mentioned code. Since the encrypted message is not readable by human eyes, it can be openly transferred to a person who is authorized to read it. 

Decrypting is a process of transferring an encrypted message back into a readable form. If the same secret code is used for both processes, the method is called symmetric cryptography. Definitely, the decrypted message should better be identical to the original one, otherwise both processes of encrypting and decrypting defeat the whole purpose of presenting the trustworthy message. 

Cryptanalysis, or trying to break a code, is the process of attempting to transfer an encrypted message into the original form without using a secret code \cite{ros} \cite{kon}. Usually, an unauthorized person may try to do it. The harder the process of breaking the code, the more secure the encrypting method. A brute force attack method is described in this work.


\section{Encrypting a Message}

Let us talk about how an English message should be written in order to be encrypted properly with the method introduced in this work. To keep it simple, we will capitalize and use all twenty-six letters of the English alphabet. Also, we will use all ten numerical digits, and four additional symbols: a period, a comma, a question mark, and an underscore. The last one will substitute a space that we normally use in between words. Thus, the first couple lines of one of the most well-known songs in English will look like this: 

\begin{equation}
O\_BEAUTIFUL\_FOR\_SPACIOUS\_SKIES,\_FOR\_AMBER\_WAVES\_OF\_GRAIN. \label{init_srting}
\end{equation}

Consequently, any message that we may want to encrypt \cite{ros} \cite{kon} \cite{mil} \cite{sin} can contain symbols from the following set:

\begin{equation}
?\_.,ABCDEFGHIJKLMNOPQRSTUVWXYZ1234567890 \label{alpha_srting}
\end{equation}

As we can see, the set contains forty symbols. Each symbol will be represented by an integer from $1$ to $40$ according to its order. The integer will be used in the encrypting and decrypting processes later in this work.

A cryptographic hash function \cite{sti}, where $y$ is an encrypted message, $h$ is a hash function, $c$ is a secret code, and $x$ is an initial message, will be used at the beginning portion of our encryption process. 

\begin{equation}
y = h_c(x) \label{h_formula}
\end{equation}

For our particular example, let us pick a ten-digit secret code $c = 0135792468$, containing only numerical digits from $0$ to $9$. This code will be used by us as a secret key to our message. 

For the first step, we will check if the code chosen by us is good enough for one of the functions that will be used later. Let us check if the number $c$ belongs to the domain $[1000000000, 9000000000]$. It does not. For this reason, we will do the following steps: \\

1. If the first digit of the code is $0$, we will substitute it with $1$ to make the whole number bigger.  

2. If the first digit is $9$ and the rest are not all zeros, we will substitute $9$ with $8$ to make the whole number smaller. \\

Thus, the code $0135792468$ will be substituted with the code $1135792468$. The reason for doing so will be explained later. Then, we will transfer the new code $1135792468$ into the decimal form $0.1135792468$ and add $1$ to it. From now on, we will use both numbers: the derived one, $a = 1.1135792468$ and the initial one, $c = 0135792468$.

Next, let us introduce the hash function that will be used for the encrypting. Let $A_n$ be an encrypted outcome from one symbol taken from the original message. Then, 

\begin{equation}
A_n = \frac{a^n \cdot 10^{20}}{\sum^{40}_{i=1} a^i}\label{1_formula}
\end{equation}
where $a = 1.1135792468$ in our particular example and $n$ is the order of a symbol to be encrypted in the set (~\ref{alpha_srting}).

Thus, $n$ is an integer, such that $1 \leq n \leq 40$ and $A_n$ is a decimal. However, we will keep only the first fifteen digits of the number $A_n$ because we want to keep it short. That is why $A_n$ will be transferred into a fifteen-digit integer $E_n$.

The reason for transferring the initial code $c = 0135792468$ into the code $1135792468$, so it would belong to the domain $[1000000000, 9000000000]$ is that the decimal $1.0135792468$ is too close to $1$. We do not want to keep it too close to $1$ because it would make all the $A_n$ too close to each other due to the way the (~\ref{1_formula}) formula is designed. That is why we are using the decimal $1.1135792468$ instead.

On the other hand, if the initial code was the number $9135792468$, we would want to transfer it into the number $8135792468$. The reason for doing so is because the decimal $1.9135792468$ is too close to $2$ and it would make all the $A_n$ values too small, again due to the way the (~\ref{1_formula}) formula is designed.

The reason for choosing the previously mentioned formula for this work is that it accepts two arguments: $a$ and $n$. The argument $n$ should better be an integer because it plays the role of an exponent in the formula and we do not want to deal with decimal exponents. However, the argument $a$ can be an integer or a lengthy decimal as well because it is a base of a power in the formula. Any minor change of the base will change the outcome, which is a good thing because we want a different result for a different secret code. Also, there is no way to figure out two initial arguments just by knowing the outcome, which is a good thing too because it makes it harder to break the code. Moreover, multiplying by $10^{20}$ makes the initial $A_n$ big enough, so if we take only the first fifteen digits and leave the rest out, we do not have to deal with a decimal point. Thus, as it was mentioned before, the final $E_n$ value will be a fifteen-digit integer.

Let us talk about the second formula that will be used by us. We will add up all the digits of the initial secret code 
$c = 0135792468$. Thus, $0 + 1 + 3 + 5 + 7 + 9 + 2 + 4 + 6 + 8 = 45$. Both digits of the sum will be used for the first and second members of a sequence which is similar to the Fibonacci one. Thus, $B_1 = 4$ and $B_2 = 5$. There are three major points of difference between the regular Fibonacci sequence and the one we will be using: \\

1. The regular one has either $F_1 = 0$ and $F_2 = 1$ or $F_1 = 1$ and $F_2 = 1$. However, as it was mentioned above, $B_1 = 4$ and $B_2 = 5$ in our particular example. If in a different example, the sum of all the digits of the number $c$ has more than two digits, the one before the last one will be assigned to $B_1$ and the last one will be assigned to $B_2$. If the sum has only one digit, its value will be assigned to $B_2$ and zero will be assigned to $B_1$. Thus, $B_1$ and $B_2$ are integers from $0$ to $9$. 

2.  The regular Fibonacci sequence's formula is $F_n = F_{n-1} + F_{n-2}$ \cite{ros}. However, we will be using the initial formula 
\begin{equation}
B_n = (B_{n-1} + B_{n-2}) \thinspace mod \thinspace 10 \label{2_formula}
\end{equation}
where $mod$ is short for modulo or modulus which is the remainder after division. Basically, we will not allow our sequence to grow. Instead, we will truncate each member which becomes greater than $9$ by leaving the last digit of it only.

3. We will not allow our sequence's members to be too small either. It will not contain any digit that is smaller than $5$. Thus, every $B_n$ that is smaller than $5$ will be turned into $C_n = 9 - B_n$ to make it greater or equal to $5$. In a case when $B_n$ is greater or equal to $5$, $C_n = B_n$. Consequently, the value of $C_n$ can be calculated with a formula 
\begin{equation}
C_n = \left\{ 
\begin{array}{ll} 
B_n & \hspace{5mm} if \thinspace B_n \geq 5 \\
9 - B_n & \hspace{5mm} if \thinspace B_n < 5 \\
\end{array} \right. 
\label{3_formula}
\end{equation} \\

Let us talk about how the sequence's members can be used for our encrypting. First, we will locate each symbol of the original message in a line one by one. Then, we will transfer every symbol into a corresponding fifteen-digit integer and we already know how to do it. Thus, each symbol will be substituted by a corresponding group of fifteen digits. Then, we will stretch the line of integers by allocating an empty spot in between every two consecutive integers. Each spot will be taken by a group of randomly generated digits. Each digit will be taken from an interval $[0, 9]$. Their quantity will be defined by the sequence in the way described below.

Each odd-indexed (first, third, fifth, and so on) member of the sequence will tell us how many digits out of each initial integer, that corresponds to one symbol, we will keep and each even-indexed member will tell us how many randomly generated misleading digits will replace each empty spot in between truncated integers. Thus, if $k$ is the length of the initial message in English that is supposed to be encrypted, we need to create $2k$ members of our sequence.

To make the whole process more understandable to the reader, let us perform all the calculations for the first three symbols of the message  (~\ref{init_srting}). The first three symbols are ``$O\_B$'' and we will encrypt them step by step using an Excel Spreadsheet for our calculations.

Let us assign a corresponding integer to each symbol according to its order in a set (~\ref{alpha_srting}). \\

``$O$'' - $19$

``$\_$'' - $2$

``$B$'' - $6$ \\

Thus, we will calculate three different $A_n$ values: $A_{19}$, $A_2$, and $A_6$. 

Let us also recall that we are using the (~\ref{1_formula}) formula, the initial secret code $c = 0135792468$ and the value $a = 1.1135792468$ in our example. \\

$A_{19} = \frac{1.1135792468^{19} \cdot 10^{20}}{\sum^{40}_{i=1} 1.1135792468^i} = \frac {7.72149587529851 \cdot 10^{20}}{715.077530341803} = 1.07981240462243 \cdot 10^{18};$ \\

$A_{2} = \frac{1.1135792468^{2} \cdot 10^{20}}{\sum^{40}_{i=1} 1.1135792468^i} = \frac {1.24005873890366 \cdot 10^{20}}{715.077530341803} = 1.73415984461281 \cdot 10^{17};$ \\

$A_{6} = \frac{1.1135792468^{6} \cdot 10^{20}}{\sum^{40}_{i=1} 1.1135792468^i} = \frac {1.90689496364995 \cdot 10^{20}}{715.077530341803} = 2.66669680242709 \cdot 10^{17};$ \\

Let us keep the first fifteen digits for each number and drop the rest. Thus, \\

$E_{19} = 107981240462243;$ 
 
$E_{2} = 173415984461281;$

$E_{6} = 266669680242709;$ \\

We have three integers now. It means that we need to calculate the first six members of our sequence. Having $B_1 = 4$ and $B_2 = 5$, let us get the rest. Using the formula $B_n = (B_{n-1} + B_{n-2})  \thinspace mod \thinspace 10$, we get: \\

$B_3 = (B_2 + B_1) \thinspace mod \thinspace 10 = (5 + 4) \thinspace mod \thinspace 10 = 9  \thinspace mod \thinspace 10 = 9;$

$B_4 = (B_3 + B_2) \thinspace mod \thinspace 10 = (9 + 5) \thinspace mod \thinspace 10 = 14 \thinspace mod \thinspace 10 = 4;$

$B_5 = (B_4 + B_3) \thinspace mod \thinspace 10 = (4 + 9) \thinspace mod \thinspace 10 = 13 \thinspace mod \thinspace 10 = 3;$

$B_6 = (B_5 + B_4) \thinspace mod \thinspace 10 = (3 + 4) \thinspace mod \thinspace 10 = 7 \thinspace mod \thinspace 10 = 7;$ \\

Let us use the rule that helps to make sure that all the members of our sequence are greater or equal to $5$. \\

$C_1 = 9 - 4 = 5$;

$C_2 = 5$;

$C_3 = 9$;

$C_4 = 9 - 4 = 5$;

$C_5 = 9 - 3 = 6$;

$C_6 = 7$; \\

The six members are: $5$, $5$, $9$, $5$, $6$, $7$. Let us form the final encrypted message. According to our encrypting rules, the final message will contain twice as many groups as the number of symbols that the initial message has. In other words, the number of groups will be equal to the number of the members in the $C$ sequence. As it was already mentioned above, each odd-indexed member will tell us the number of digits that we will keep out of the initial integer's digits and each even-indexed member will tell us the number of randomly generated misleading digits in a group of digits that will replace each empty spot in between truncated integers.

First, our C++ program will generate a random number from $1$ to $7$. Let us suppose, it generated the number $3$. Since $C_1 = 5$, we will pick five digits from $E_{19}$ starting at the third position according to the randomly generated number. The string will be ``$79812$''. Since $C_2 = 5$, our program will generate five random numbers from $0$ to $9$ and locate them right after the sequence ``$79812$''. Let us suppose that the generated numbers are ``$95916$'' and that is why the new string is ``$7981295916$''. 

Then, a random number from $1$ to $7$ will be generated again. Let us suppose it is the number $5$. Since $C_3 = 9$, we will keep nine digits from the $E_2$ number starting at the fifth position and attach them to our string. Thus, attaching ``$159844612$'' to ``$7981295916$'' will give us ``$7981295916159844612$''. Since $C_4 = 5$, the program will generate five numbers from $0$ to $9$. Let us suppose they are ``$33613$'', so the new sequence is ``$798129591615984461233613$''. 

One more random number from $1$ to $7$ will be generated and let us suppose it was the number $7$. Since $C_5 = 6$, we will pick six digits of the $E_6$ value starting at the seventh position. They are ``$680242$'' and that is why the new string is ``$798129591615984461233613680242$''. Finally, $C_6 = 7$. It means that the C++ program will generate seven numbers from $0$ to $9$. Let us suppose they are ``$3427975$''. Thus, the new string is ``$7981295916159844612336136802423427975$''.

The whole process of attaching truncated $E_n$ values and groups of random digits will continue until we reach the very end of our initial message that we are trying to encrypt. After the last $E_n$ value gets truncated and attached, the last group of random digits gets generated and attached as well. Then the process is finished and the final encrypted message is created.

The encrypting C++ program, written for this purpose and presented in the Appendix ~\ref{Enc}, will read information from two files. One file will submit an initial message and the other one will submit the secret code. Then, a file containing the encrypted message is created.

We should realize that every time we run the C++ program with the same secret code and the same initial message, we will get a different final message due to that fact that we are using random numbers. Thus, if we run the program using a secret code $c = 0135792468$ and the full initial message (~\ref{init_srting}), we might get a string that looks like this: \\ \\
$1079871165598446128842982666697195232245141846990925370770494690888836061815282566663925014862$ \\
$6737411811075007010894420888360619439159781960699655547936734159844017451063147695215736625240$ \\
$4642510956919426388092395341598901347685602519079002456684528774707704983406309578211644886626$ \\
$7374593911812405090583260590880490188768563752322415984438491949847685602405948213048063754083$ \\
$6243673745587493068245497822716047680537323404600207449844612812814470147609488468124046221721$ \\
$9011169194192727341598439300770498074719135206646537456792666648822614184652005914451116919422$ \\
$1001037641598446128987001953391105259280394707509388322929587834342102451410528904768560565680$ \\
$1734159681390981240462521910701476066898598446128052446456071046334269194294899449470736978493$ \\
$745572988374425719874933487604003112441377570298$ \\

As we can see, the beginning of the string above does not look exactly like the beginning of the string ``$7981295916159844612336136802423427975$'', however the substring ``$798$'' can be found at the beginning of both of them because this substring is a portion of a bigger string that was initiated by the same symbol.

Thus, we got a lengthy string, containing exactly eight hundred digits, which was derived from our initial message and the secret ten-digit code. This sequence can be sent as an encrypted message to a person who is authorized to read it. We assume that the person already has the secret code.


\section{Decrypting a Message}

Let us talk about decrypting the lengthy message containing eight hundred digits described above. We will have to use all the steps mentioned in the first chapter, but in the reverse order \cite{ros} \cite{sin}.

First, we will recall that the initial secret code is $c = 0135792468$ and the sum of its digits is $45$. The two digits $4$ and $5$ will produce the sequence mentioned above. The C++ program that was written specifically to decrypt any message by this method will figure out how many members of this sequence are needed. We know that we are decrypting a message that is eight hundred symbols long. Thus, the program will calculate the members of the sequence and their intermediate totals. As soon as the total reaches $800$, the process will stop.

Let us do a couple of steps manually, just using the Excel spreadsheet; however, we still want to make it easier for us. Thus, we will not calculate that many members of the sequence and all their intermediate totals. Instead, we will calculate six consecutive members starting from the seventh one because the first six members were already calculated in the first chapter. This is how we will decrypt the fourth, fifth, and sixth symbol of the message.

Let us restate what we have calculated before. \\

$B_5 = 3$;

$B_6 = 7$; \\

Using the $B_n$ formula, we get: \\

$B_7 = (B_6 + B_5) \thinspace mod \thinspace 10 = (7 + 3) \thinspace mod \thinspace 10 = 10 \thinspace mod \thinspace 10 = 0$;

$B_8 = (B_7 + B_6) \thinspace mod \thinspace 10 = (0 + 7) \thinspace mod \thinspace 10 = 7 \thinspace mod \thinspace 10 = 7$;

$B_9 = (B_8 + B_7) \thinspace mod \thinspace 10 = (7 + 0) \thinspace mod \thinspace 10 = 7 \thinspace mod \thinspace 10 = 7$;

$B_{10} = (B_9 + B_8) \thinspace mod \thinspace 10 = (7 + 7) \thinspace mod \thinspace 10 = 14 \thinspace mod \thinspace 10 = 4$;

$B_{11} = (B_{10} + B_9) \thinspace mod \thinspace 10 = (4 + 7) \thinspace mod \thinspace 10 = 11 \thinspace mod \thinspace 10 = 1$;

$B_{12} = (B_{11} + B_{10}) \thinspace mod \thinspace 10 = (1 + 4) \thinspace mod \thinspace 10 = 5 \thinspace mod \thinspace 10 = 5$; \\

Then, using the $C_n$ formula, we get: \\

$C_7 = 9 - 0 = 9$;

$C_8 = 7$;

$C_9 = 7$;

$C_{10} = 9 - 4 = 5$;

$C_{11} = 9 - 1 = 8$;

$C_{12} = 5$; \\

Now, we need to restate all the previous $C_n$ members. \\

$C_1 = 5$;

$C_2 = 5$;

$C_3 = 9$;

$C_4 = 5$;

$C_5 = 6$;

$C_6 = 7$; \\

Since we are not planning to decrypt the first three symbols, it would be enough if we just get the total of all the $C_n$ values listed above. \\

$C_1 + C_2 + C_3 + C_4 + C_5 + C_6 = 5 + 5 + 9 + 5 + 6 + 7 = 37$; \\

Thus, starting at the $38th$ position of the encrypted message above, we will pick groups of $9$, $7$, $7$, $5$, $8$, and $5$ digits. Let us list the groups: \\

``$245141846$''

``$9909253$'' 

``$7077049$'' 

``$46908$'' 

``$88836061$''

``$81528$'' \\

The second, fourth, and sixth groups have no meaning for us because they are the groups of misleading digits. Thus, we can ignore them and leave just the first, third, and fifth groups because they represent the fourth, fifth, and sixth symbols of the message that we are trying to encrypt. The groups that we need are: \\

``$245141846$''

``$7077049$''

``$88836061$'' \\

The easier part of the decrypting process is done. Now, we will do the harder one. Even though we are trying to decrypt only three symbols from the encrypted message, we will have to calculate all forty $E_n$ values simply because we do not know what symbols we are getting before we decrypt them. In fact, we actually do know them because they are the fourth, fifth, and sixth symbols of the initial message, but let us suppose that we do not, since the person who will decrypt the message will initially not know any symbols from it.

Let us use the Excel spreadsheet and calculate the $A_n$ and $E_n$ values. We will print the final $E_n$ values that have fifteen digits each. \\

$E_{1 } = 155728462935720$

$E_{2 } = 173415984461281$

$E_{3 } = 193112441359474$

$E_{4 } = 215046006996792$

$E_{5 } = 239470770498836$

$E_{6 } = 266669680242709$

$E_{7 } = 296957821669073$

$E_{8 } = 330686067385615$

$E_{9 } = 368245141846527$

$E_{10 } = 410070147695215$

$E_{11 } = 456645606205602$

$E_{12 } = 508511070212964$

$E_{13 } = 566267374557214$

$E_{14 } = 630583596446836$

$E_{15 } = 702204806375703$

$E_{16 } = 781960699383195$

$E_{17 } = 870775206646340$

$E_{18 } = 969677198749346$

$E_{19 } = 107981240462243$

$E_{20 } = 120245668422474$

$E_{21 } = 133903080872861$

$E_{22 } = 149111691942601$

$E_{23 } = 166047685602515$

$E_{24 } = 184907256666132$

$E_{25 } = 205908883606125$

$E_{26 } = 229295859515538$

$E_{27 } = 255339110533671$

$E_{28 } = 284340334386668$

$E_{29 } = 316635495401165$

$E_{30 } = 352598716478975$

$E_{31 } = 392646613119303$

$E_{32 } = 437243119695965$

$E_{33 } = 486904863899515$

$E_{34 } = 542207151604478$

$E_{35 } = 603790631493288$

$E_{36 } = 672368716643193$

$E_{37 } = 748735849051409$

$E_{38 } = 833776702838826$

$E_{39 } = 928476432746648$

$E_{40 } = 103393208664956$ \\

If we use a search tool to look for the three groups of digits ``$245141846$'', ``$7077049$'' and ``$88836061$'' among the above listed $E_n$ values, we would be able to find out that ``$245141846$'' is a substring of ``$368245141846527$'' which equals to $E_9$. Then, ``$7077049$'' is a substring of ``$239470770498836$'' which equals to $E_5$. Finally, ``$88836061$'' is a substring of ``$205908883606125$'' which equals to $E_{25}$. 

Thus, the fourth, fifth, and sixth symbols of the message correspond to the indexes $9$, $5$, and $25$. If we go back to the set (~\ref{alpha_srting}) , we will see that the ninth symbol of it is ``$E$'', the fifth is ``$A$'', and the twenty fifth is ``$U$''. Consequently, the symbols that we were looking for are ``$EAU$''. 

Looking at the initial message (~\ref{init_srting}) that was encrypted, we can see that the symbols were decrypted correctly.

Just like the encrypting program, the C++ decrypting program, written for this section and presented in the  Appendix ~\ref{Dec}, will read information from two files as well. One file will submit a lengthy encrypted message and the other one will submit the secret code. Then, the decrypted message will be printed into an output file. Definitely, the program will decrypt the whole entire message much faster than we did for just three symbols manually.

Please notice that each substring in our example was found exactly once. The chances that at least one substring can be found more than once are very low due to the uniqueness of the $A_n$ values. However, the C++ program takes care of this rare ``double-matching'' situation as well. In the case when at least one string was found more than once, the decrypted message will not be printed. A statement ``The initial encoded message needs to be resent'' will be printed instead. If this happens, the initial message in English will be encoded again by a person who is supposed to do it with the same secret code that was used the first time. As it was mentioned before, the message encoded the second time will look different from the first one due to it inserting random numbers into the encrypting process.  

One more unusual situation may happen when the message is being decrypted. If a wrong secret code is mistakenly used, then the encrypted message will not be decrypted at all. A message ``An incorrect code was used for decoding.''  will be printed instead.


\section{Breaking the Code}

We had a chance to talk about how to encrypt and decrypt a message using a secret code. The initial message that two parties are hiding from others is not intended to be read by anybody else. Unfortunately, there is a chance that another person or a group of people will make an attempt to break the code with a purpose of reading the message. The goal of the person who is designing the encrypting method is to make the task of breaking the code as hard as possible \cite{ros} \cite{kon}.

In this work, we make an assumption that a party trying to break our code knows the method used to encrypt and decrypt a message, but does not know the actual secret code. In reality, if a person trying to break this code is not familiar with the method we are using, the chances to succeed are equal to zero. Thus, let us imagine the worst case scenario: the person not only knows the method, but also has access to both C++ program that we are using for encrypting and decrypting. The only piece of information that must stay hidden at all times is the secret code, otherwise the message could be decrypted easily and there would be nothing to break after that.

Let us recall that the decrypting C++ program reads information from two files. One file submits an encrypted message and the other one submits the secret code. Let us imagine that a person, who has access to the decrypting program and the encrypted message, wants to break the code and read the message. An attempt to achieve this goal can be done if the person modifies the program in a particular way. 

Due to that fact that the secret code is not known by the person, a program will not read the code from a file, but should try to guess it instead. Since the code is a ten-digit number, it might take up to $10^{10}$ attempts to guess it. If a loop with $10^{10}$ iterations gets inserted into a C++ program, it will freeze. This is how a person who is trying to break the code will face his first obstacle: the loop trying to guess the code cannot be inserted into the program as one piece. It needs to be broken into smaller ones.

The C++ program supporting this work can contain a loop that attempts to guess the code with $100$ iterations easily. Then, if we use a larger range for the loop, the program might freeze due to the fact that each single iteration already represents a relatively long process. 

If we pick and use a relatively narrow range that includes the secret code, the program will guess the code correctly. Then, it will print the decrypted message and the secret code into a file. A message ``The secret code was determined successfully. The printed message has no known issues.'' will be printed as well.
For instance, the secret code $c = 0135792468$ was used in our example in previous chapters. If we pick the range from $0135792400$ to $0135792500$, we will succeed in breaking the code.

In the case when the decrypting process faces the ``double-matching'' issue described above, the decrypted message in this situation will be printed anyway. Then the message ``The program was able to determine the secret code; however, the printed message might have some issues.'' will be printed as well as the hypothetical secret code. The reason why we will treat the same situation differently in the previous chapter is because a person who has the secret code can request an encrypted message to be resent again and then that message will be decrypted easily with the code. However, the person breaking the code cannot ask to resend it. That is why whatever was decrypted by breaking the code is valuable in any situation. 

Once again, a person trying to break the code does not know it, that is why the best range cannot be chosen immediately. It needs to be found. Dividing $10^{10}$ by $100$ gives us $10^8$ attempts to guess the correct range. This is how the person breaking the code will face yet one more obstacle: trying $10^8$ loops one after another is not a quick task. Every time a wrong range gets chosen, the message ``The program was not able to determine the secret code successfully. Try a different range.'' will be printed.  The C++ breaking the code program is presented in the  Appendix ~\ref{Br}.


\section{Conclusion}

We have discussed a method to encrypt a message, a method to decrypt a message, and a method to break the code. All three processes have their own goals and requirements. Since we already had a chance to talk about what all three of them are supposed to do, let us talk about the requirements they have to meet \cite{kon} \cite{mil}.

The encrypting process must create an encrypted message that cannot be read by the human eye. Also, if the message can be decrypted easily without even using a computer, the method is worthless. On the other side, one might argue that such a short initial message  (~\ref{init_srting}) should not be producing an outcome containing eight hundred digits. However, we might argue back saying that this long encrypted message should not bother the C++ program which will decrypt it in seconds anyway regardless of its length.

Talking about the decrypting process, we definitely should mention that it must be authentic. In other words, the decrypted message must be absolutely the same as the initial message, otherwise the method is useless. Thus, if the ``double-matching'' issue takes place in a decrypting process, the program will not print the questionable decrypted message, but will print an error message instead.

The only reason why we are attempting to break the code in this work is because we are trying to see how to make the code as unbreakable as possible. We already found out that it might take up to $10^8$ loops to break the code. Let us think about increasing the security of the code even more. 

A ten-digit secret code was chosen because the C++ program that was used to support this work has a hard time holding values of variables longer than that. However, the program could be designed differently. If we wanted to use a secret code containing $30$ digits, we could use three different variables one by one storing the secret code's value. In this case, the program would have to be changed to be much more complex. However, it is doable and worth it. 

If a longer secret code is used, the encrypting and decrypting processes will take more time. It means that one iteration of attempting to break the code becomes longer as well. Most definitely, one loop attempting to guess the secret code will not be able to contain $100$ iterations anymore, but will be able to hold a smaller number of iterations instead. On the other hand, a thirty-digit code can be selected out of $10^{30}$ choices. Dividing this number by the number of loop's iterations that is less than $100$, we get that the number of loops will be greater than $10^{28}$. Thus, using a group of variables to store the secret code's value would make the guessing process tedious and worthless.

Let us also recall that the whole idea of an unauthorized person trying to modify the existing C++ decrypting program with the purpose of breaking the code is based on our assumption that the person has access to the program. In reality, if a person does not know the method we are using, there is no chance to succeed.

Finally, let us discuss the main formula that was used for encrypting and the reason why it was selected by the author. According to the encrypting design of this work, a message containing characters from the set (~\ref{alpha_srting}) should be encrypted character by a character. Since, the set contains forty characters, one argument of the formula should take a positive integer from $1$ to $40$ to represent a character to be encrypted. Next, since a ten-digit secret code is used for encrypting as well, another argument should take care of the long number. Theoretically, we could leave the number as an integer. However, using such a big value in a formula might be uncomfortable. That is why the decision of converting the integer into a decimal by attaching a ``1.'' string in front of the integer was made.

Thus, we have to pick a formula that takes two arguments. Moreover, one of the arguments should be able to accept a lengthy decimal. The formula (~\ref{1_formula}) is a modified version of the formula 

\begin{equation}
P_k = \frac{a^k }{\sum^9_{i=1} a^i}\label{prob_formula}
\end{equation}
that was derived by the same author \cite{pas} in an article talking about first digit's probabilities. In this formula, $P_k$ is the probability that the value of the $f(x) = log_a(x)$ function will start with the digit $k$ having the range $[1, 10)$. The reason why the formula  (~\ref{prob_formula}) was used as the base for the modified one is because it takes two arguments $a$ and $k$. In addition, we have to choose which one of them is more suitable for taking a decimal value. According to the formula's structure, $a$ is a base of a power, but $k$ is an exponent. That is why, it looks like the argument $a$ can accept a decimal value as well as an integer, while the argument $k$ should accept integers only if we are trying to keep our results neat.

Now, we are ready to talk about why the selected (~\ref{prob_formula}) formula had to be modified. To have a better chance to operate with all the $A_n$ values from $A_1$ to $A_{40}$, they should initially belong to the same uniform ranges. To be more specific, it would be nice if all of them were either less or greater than $1$. However, allowing them to be initially greater than $1$ means that their common domain will be unbounded at one end. Since we would prefer to keep all the situations in this work as predictable as possible, let us keep all the $A_n$ values in the initial range $(0, 1)$ because it is bounded at both ends. This goal can be accomplished by modifying the sum in the  (~\ref{prob_formula}) formula. Adding up all forty $a^i$ values instead of just nine of them will make sure that a fraction $\frac{a^n}{\sum^{40}_{i=1} a^i}$ will always produce a value that is less than $1$. We just need to make sure that $n$ is an integer from $1$ to $40$.

Now, when we have all $A_n$ values in the initial range $(0, 1)$, we realize that it is easier to deal with integers then with decimals; especially, if we are planning to substitute each character with a string of digits, we do not need any decimal point inside of a string. That is why, every $\frac{a^n}{\sum^{40}_{i=1} a^i}$ value will be multiplied by $10^{20}$. Thus, we are getting the (~\ref{1_formula}) formula.

There is a chance that someone would want to modify the initial formula (~\ref{prob_formula}) differently. Moreover, another person might want to use a different formula as the main one and/or substitute the modified Fibonacci (~\ref{2_formula}) formula with another one. Furthermore, someone else might offer just one fabulous formula that will do the whole job all at once. The purpose of this work is to show how any applicable formula or a group of such formulas can be used to accomplish the task of encrypting and decrypting.


\section{Acknowledgements}
I would like to thank my son, Vitaliy Goncharenko, for his emotional support.

\appendix
\section{C++ Programs} 
\subsection{Encrypting Program} \label{Enc}

\vspace{10pt} 

\noindent
\#include $<$iostream$>$ 

\noindent
\#include $<$fstream$>$ 

\noindent
\#include $<$string$>$ 

\noindent
\#include $<$stdio.h$>$

\noindent
\#include $<$stdlib.h$>$ // atof 

\noindent
\#include $<$sstream$>$ 

\noindent
\#include $<$cmath$>$ 

\noindent
\#include $<$time.h$>$  \\

\noindent
using namespace std;  \\

\noindent
int main() \{

	string secretCode;  //reads from a file

	string secretMessage;  //reads from a file

	string stringCodeItem;  //function value for an individual letter

	string stringTempMisleadItem;   //temporary storage for misleading item

	string stringFinalCodedMessage;  //final encrypted message printed into output file

	const string alphanumeric = "?\_.,ABCDEFGHIJKLMNOPQRSTUVWXYZ1234567890"; \\

	char charAlphaNumeric[41];  //alphabet items

	char charInitMessage[101];  //initial message items

	char charCodeItem[41][16];  //function value for an individual letter

	char charRandomCodeItem[101][10];  //truncated function value for an individual letter

	char charRandomMisleadItem[101][10];  //group of random misleading digits \\

	int codeLength;  //lenght of the secretCode

	int messageLenght;  //lenght of the secretMessage

	int codeSum;  //sum of the code digits 
	int firstFib;  //first number to start Fib

	int secondFib;  //second number to start Fib

	int randomValue;  //random value \\

	int initialCode[11];  //array for initial numerical secret code

	int completedFullCode[201];  //array for completed full numerical secret code

	int completedOddFullCode[101];  //array for completed odd full numerical secret code

	int completedEvenFullCode[101];  //array for completed even full numerical secret code \\

	long double numberCode;  //numerical secret code

	long double modifiedCode;  //numerical secret code with a shorter range

	long double sumItem;  //part of the major function

	long double diffValue;  //part of the initialCode forming function \\

	long double codeItem[41];  //function value for an individual letter \\

	ifstream codeFile;

	ifstream messageFile; \\ 

	ofstream outputFile; \\

	srand(time(NULL)); \\

	//initializing all the strings

	secretCode = "";   

	secretMessage = "";

	stringCodeItem = "";

	stringTempMisleadItem = "";

	stringFinalCodedMessage = ""; \\

	//initializing all the numbers

	codeLength = 0;  

	messageLenght = 0;

	codeSum = 0;

	firstFib = 0;

	secondFib = 0;

	randomValue = 0; \\

	numberCode = 0.0;

	modifiedCode = 0.0;

	sumItem = 0.0;

	diffValue = 0.0; \\

	int k = 0;

	int j = 0;

	int n = 0; \\

	//initializing all the arrays

	for (k = 0; k $\leq$ 10; k++)  

	\{
	
		\qquad initialCode[k] = 0;

	\} \\

	for (k = 0; k $\leq$ 40; k++)

	\{

		\qquad codeItem[k] = 0;

		\qquad charAlphaNumeric[k] = ' '; \\

		\qquad for (j = 0; j $\leq$ 15; j++)

		\qquad \{

			\qquad \qquad charCodeItem[k][j] = ' ';

		\qquad \}

	\} \\

	for (k = 0; k $\leq$ 100; k++)

	\{

		\qquad charInitMessage[k] = ' ';

		\qquad completedOddFullCode[k] = 0;

		\qquad completedEvenFullCode[k] = 0;

	\} \\

	for (k = 0; k $\leq$ 200; k++)

	\{

		\qquad completedFullCode[k] = 0;

	\} \\

	for (k = 0; k $\leq$ 100; k++)

	\{

		\qquad for (j = 0; j $\leq$ 9; j++)

		\qquad \{

			\qquad \qquad charRandomCodeItem[k][j] = ' ';

			\qquad \qquad charRandomMisleadItem[k][j] = ' ';

		\qquad \}

	\} \\

	codeFile.open("Code.txt");  

	messageFile.open("Initial\_Message.txt"); \\

	outputFile.open("Final\_Coded\_Message.txt"); \\

	getline(codeFile, secretCode); 

	getline(messageFile, secretMessage); \\

	codeLength = secretCode.length();

	messageLenght = secretMessage.length(); \\

	//forming function value for an individual letter

	for (k = 1; k $\leq$ 40; k++)  

	\{

		\qquad charAlphaNumeric[k] = alphanumeric[k - 1];

	\} \\

	for (k = 1; k $\leq$ messageLenght; k++)

	\{

		\qquad charInitMessage[k] = secretMessage[k - 1];

	\} \\

	numberCode = atof(secretCode.c\_str()); \\

	if (numberCode $<$ 1 * pow(10, codeLength - 1))

	\{

		\qquad modifiedCode = numberCode + pow(10, codeLength - 1);

	\}

	else

	\{

		\qquad if (numberCode $>$ 9 * pow(10, codeLength - 1))

		\qquad \{

			\qquad \qquad modifiedCode = numberCode - pow(10, codeLength - 1);

		\qquad \}

		\qquad else

		\qquad \{

			\qquad \qquad modifiedCode = numberCode;

		\qquad \}

	\} \\

	modifiedCode = modifiedCode / pow(10, codeLength) + 1; \\

	for (k = 1; k $\leq$ 40; k++)

	\{

		\qquad sumItem = sumItem + pow(modifiedCode, k);

	\} \\

	for (k = 1; k $\leq$ 40; k++)

	\{

		\qquad codeItem[k] = pow(modifiedCode, k) * pow(10, 30) / sumItem;

		\qquad stringCodeItem = to\_string(codeItem[k]); \\

		\qquad for (j = 1; j $\leq$ 15; j++)

		\qquad \{

			\qquad \qquad charCodeItem[k][j] = stringCodeItem[j - 1];

		\qquad \}

	\} \\

	//forming completedOddFullCode and completedEvenFullCode

	for (k = 1; k $\leq$ codeLength; k++)  

	\{

		\qquad initialCode[k] = numberCode / pow(10, codeLength - k) - diffValue;

		\qquad diffValue = (diffValue + initialCode[k]) * 10;

		\qquad codeSum = codeSum + initialCode[k];

	\} \\

	firstFib = (codeSum \% 100) / 10;

	secondFib = codeSum \% 10; \\

	completedFullCode[1] = firstFib;

	completedFullCode[2] = secondFib; \\

	for (k = 3; k $\leq$ messageLenght * 2; k++)

	\{

		\qquad completedFullCode[k] = (completedFullCode[k - 1] + completedFullCode[k - 2]) \% 10;

	\} \\

	for (k = 1; k $\leq$ messageLenght * 2; k++)

	\{

		\qquad if (completedFullCode[k] $<$ 5)

		\qquad \{

			\qquad \qquad completedFullCode[k] = 9 - completedFullCode[k];

		\qquad \}

	\} \\

	for (k = 1; k $\leq$ messageLenght; k++)

	\{

			\qquad completedOddFullCode[k] = completedFullCode[k * 2 - 1];

			\qquad completedEvenFullCode[k] = completedFullCode[k * 2];

	\} \\

	// forming an array of meaningful code

	for (k = 1; k $\leq$ messageLenght; k++) 

	\{

		\qquad for (j = 1; j $\leq$ 40; j++)

		\qquad \{

			\qquad \qquad // finding the matching letter in the alphabet

			\qquad \qquad if (charInitMessage[k] == charAlphaNumeric[j]) 

			\qquad \qquad \{

				\qquad \qquad \qquad // picking a starting point for the selected string

				\qquad \qquad \qquad randomValue = rand() \% 7 + 1;  \\

				\qquad \qquad \qquad for (n = 1; n $\leq$ completedOddFullCode[k]; n++)

				\qquad \qquad \qquad \{

					\qquad \qquad \qquad \qquad charRandomCodeItem[k][n] = charCodeItem[j][randomValue + n - 1];

				\qquad \qquad \qquad \}

			\qquad \qquad \}

		\qquad \}

	\} \\

	// forming an array of misleading code

	for (k = 1; k $\leq$ messageLenght; k++) 

	\{

		\qquad for (n = 1; n $\leq$ completedEvenFullCode[k]; n++)

		\qquad \{

			\qquad \qquad randomValue = rand() \% 10;

			\qquad \qquad stringTempMisleadItem = to\_string(randomValue);

			\qquad \qquad charRandomMisleadItem[k][n] = stringTempMisleadItem[0];

		\qquad \}

	\} \\

	// forming the final coded message printed into output file

	for (k = 1; k $\leq$ messageLenght; k++) 

	\{

		\qquad for (j = 1; j $\leq$ completedOddFullCode[k]; j++)

		\qquad \{

			\qquad \qquad stringFinalCodedMessage = stringFinalCodedMessage + charRandomCodeItem[k][j];

		\qquad \} \\

		\qquad for (n = 1; n $\leq$ completedEvenFullCode[k]; n++)

		\qquad \{

			\qquad \qquad stringFinalCodedMessage=stringFinalCodedMessage+charRandomMisleadItem[k][n];

		\qquad \}

	\} \\
	
	// printing the final coded message into output file

	outputFile $<$$<$$<$ stringFinalCodedMessage $<$$<$ endl;  \\

	codeFile.close();

	messageFile.close(); \\

	outputFile.close(); \\

	return 0; 

\noindent
\} \\

\subsection{Decrypting Program} \label{Dec}

\vspace{10pt} 

\noindent
\#include $<$iostream$>$

\noindent
\#include $<$fstream$>$

\noindent
\#include $<$string$>$

\noindent
\#include $<$stdio.h$>$

\noindent
\#include $<$stdlib.h$>$ // atof

\noindent
\#include $<$sstream$>$

\noindent
\#include $<$cmath$>$ \\

\noindent
using namespace std; \\

\noindent
int main() \{

	string secretCode;  //reads from a file

	string initialCodedMessage;  //reads from a file

	string stringCodeItem;  //temporary storage for a function value of an individual letter

	string stringTruncItem;  //temporary storage for a truncated function value of a coded message

	string stringRandomCodeItem;   //merged all truncated function values of a coded message

	string stringFinalDecodedMessage;  //final decoded message printed into output file

	const string alphanumeric = "?\_.,ABCDEFGHIJKLMNOPQRSTUVWXYZ1234567890"; \\

	char charAlphaNumeric[41];  //alphabet items

	char charFinalDecodedMessage[101];  //final decoded message items

	char charCodeItem[41][16];  //function value for an individual letter

	char charRandomCodeItem[101][10];  //truncated function value for an individual letter \\

	int codeLength;  //lenght of the secretCode

	int codedMessageLenght;  //lenght of the initialCodedMessage

	int codedShortMessageLenght;  //lenght of the stringRandomCodeItem

	int decodedMessageLength;  //lenght of the decoded message

	int codeSum;  //sum of the code digits 

	int firstFib;  //first number to start Fib

	int secondFib;  //second number to start Fib

	int extraSum;  //sum of the misleading items

	int found;   //tells at what position if any a substring was found in a string

	int messageLevel;  //tells if the message was decoded correctly  \\

	int initialCode[11];  //array for initial numerical secret code

	int completedFullCode[201];  //array for completed full numerical secret code

	int completedOddFullCode[101];  //array for completed odd full numerical secret code

	int completedEvenFullCode[101];  //array for completed even full numerical secret code

	int corrSymbolLocations[101];  //array of correct locations of symbols in the final decoded message

	int uniqueCode[101];  //tells how many times each symbol was found \\

	long double numberCode;  //numerical secret code

	long double modifiedCode;  //numerical secret code with a shorter range

	long double sumItem;  //part of the major function

	long double diffValue;  //helps to find codeSum \\

	long double codeItem[41];  //function value for an individual letter \\

	ifstream codeFile;

	ifstream initialFile; \\

	ofstream outputFile; \\

	secretCode = "";

	initialCodedMessage = "";

	stringCodeItem = "";

	stringRandomCodeItem = "";

	stringTruncItem = "";

	stringFinalDecodedMessage = ""; \\

	codeLength = 0;

	codedMessageLenght = 0;

	codedShortMessageLenght = 0;

	decodedMessageLength = 0;

	codeSum = 0;

	firstFib = 0;

	secondFib = 0;

	extraSum = 0;

	found = -1;

	messageLevel = 1; \\

	numberCode = 0.0;

	modifiedCode = 0.0;

	sumItem = 0.0;

	diffValue = 0.0; \\

	int k = 0;

	int j = 0;

	int n = 0;

	int m = 0; \\

          	//initializing all the arrays

	for (k = 0; k $\leq$ 10; k++)

	\{

		\qquad initialCode[k] = 0;

	\} \\

	for (k = 0; k $\leq$ 40; k++)

	\{

		\qquad codeItem[k] = 0;

		\qquad charAlphaNumeric[k] = ' '; \\

		\qquad for (j = 0; j $\leq$ 15; j++)

		\qquad \{

			\qquad \qquad charCodeItem[k][j] = ' ';

		\qquad \}

	\} \\

	for (k = 0; k $\leq$ 100; k++)

	\{

		\qquad completedOddFullCode[k] = 0;

		\qquad completedEvenFullCode[k] = 0;

		\qquad corrSymbolLocations[k] = 0;

		\qquad uniqueCode[k] = 0; \\

		\qquad for (j = 0; j $\leq$ 9; j++)

		\qquad \{

			\qquad \qquad charRandomCodeItem[k][j] = ' ';

		\qquad \} \\

		\qquad charFinalDecodedMessage[k] = ' ';

	\} \\

	for (k = 0; k $\leq$ 200; k++)

	\{

		\qquad completedFullCode[k] = 0;

	\} \\

	codeFile.open("Code.txt");

	initialFile.open("Initial\_Coded\_Message.txt"); \\

	outputFile.open("Final\_Decoded\_Message.txt"); \\

	getline(codeFile, secretCode);

	getline(initialFile, initialCodedMessage); \\

	codeLength = secretCode.length();

	codedMessageLenght = initialCodedMessage.length(); \\

	//forming function value for an individual letter

	for (k = 1; k $\leq$ 40; k++)  

	\{

		\qquad charAlphaNumeric[k] = alphanumeric[k - 1];

	\} \\

	numberCode = atof(secretCode.c\_str()); \\

	if (numberCode $<$ 1 * pow(10, codeLength - 1))

	\{

		\qquad modifiedCode = numberCode + pow(10, codeLength - 1);

	\}

	else

	\{

		\qquad if (numberCode $>$ 9 * pow(10, codeLength - 1))

		\qquad \{

			\qquad \qquad modifiedCode = numberCode - pow(10, codeLength - 1);

		\qquad \}

		\qquad else

		\qquad \{

			\qquad \qquad modifiedCode = numberCode;

		\qquad \}

	\} \\

	modifiedCode = modifiedCode / pow(10, codeLength) + 1; \\

	for (k = 1; k $\leq$ 40; k++)

	\{

		\qquad sumItem = sumItem + pow(modifiedCode, k);

	\} \\

	for (k = 1; k $\leq$ 40; k++)

	\{

		\qquad codeItem[k] = pow(modifiedCode, k) * pow(10, 30) / sumItem;

		\qquad stringCodeItem = to\_string(codeItem[k]); \\

		\qquad for (j = 1; j $\leq$ 15; j++)

		\qquad \{

			\qquad \qquad charCodeItem[k][j] = stringCodeItem[j - 1];

		\qquad \}

	\} \\

	 //forming completedOddFullCode and completedEvenFullCode

	for (k = 1; k $\leq$ codeLength; k++) 

	\{

		\qquad initialCode[k] = numberCode / pow(10, codeLength - k) - diffValue;

		\qquad diffValue = (diffValue + initialCode[k]) * 10;

		\qquad codeSum = codeSum + initialCode[k];

	\} \\

	firstFib = (codeSum \% 100) / 10;  

	secondFib = codeSum \% 10; \\

	completedFullCode[1] = firstFib;

	completedFullCode[2] = secondFib; \\

	for (k = 3; k $\leq$ 200; k++)

	\{

		\qquad completedFullCode[k] = (completedFullCode[k - 1] + completedFullCode[k - 2]) \% 10;

	\} \\

	for (k = 1; k $\leq$ 200; k++)

	\{

		\qquad if (completedFullCode[k] $<$ 5)

		\qquad \{

			\qquad \qquad completedFullCode[k] = 9 - completedFullCode[k];

		\qquad \}

	\} \\

	for (k = 1; k $\leq$ 100; k++)

	\{

		\qquad completedOddFullCode[k] = completedFullCode[k * 2 - 1];

		\qquad completedEvenFullCode[k] = completedFullCode[k * 2];

	\} \\

	stringRandomCodeItem = ""; \\

	for (k = 1; k $\leq$ 100; k++) //forming a string with concatenated trancated real items

	\{

		\qquad for (n = 1; n $\leq$ completedOddFullCode[k]; n++) 

		\qquad \{

			\qquad \qquad if (n + extraSum $\leq$ codedMessageLenght)

			\qquad \qquad \{

				\qquad \qquad \qquad stringRandomCodeItem=stringRandomCodeItem+initialCodedMessage[n+extraSum-1];

			\qquad \qquad \}

		\qquad \} \\

		\qquad extraSum = extraSum + completedOddFullCode[k] + completedEvenFullCode[k];

	\} \\

	codedShortMessageLenght = stringRandomCodeItem.length();

	extraSum = 0; \\

	for (k = 1; k $\leq$ 100; k++) //forming charRandomCodeItem[][]

	\{

		\qquad for (n = 1; n $\leq$ completedOddFullCode[k]; n++)

		\qquad \{

			\qquad \qquad if (n + extraSum $\leq$ codedShortMessageLenght)

			\qquad \qquad \{

				\qquad \qquad \qquad charRandomCodeItem[k][n]=stringRandomCodeItem[n+extraSum-1];

			\qquad \qquad \}

		\qquad \} \\

		\qquad if (completedOddFullCode[k] + extraSum $\leq$ codedShortMessageLenght)

		\qquad \{

			\qquad \qquad decodedMessageLength = decodedMessageLength + 1;

		\qquad \} \\

		\qquad extraSum = extraSum + completedOddFullCode[k];

	\} \\

	for (k = 1; k $\leq$ decodedMessageLength; k++)  //finding the matching symbol

	\{

		\qquad stringTruncItem = ""; \\

		\qquad for (j = 1; j $\leq$ completedOddFullCode[k]; j++)

		\qquad \{

			\qquad \qquad stringTruncItem = stringTruncItem + charRandomCodeItem[k][j];

		\qquad \} \\

		\qquad for (n = 1; n $\leq$ 40; n++)

		\qquad \{

			\qquad \qquad stringCodeItem = ""; \\

			\qquad \qquad for (m = 1; m $\leq$ 15; m++)

			\qquad \qquad \{

				\qquad \qquad \qquad stringCodeItem = stringCodeItem + charCodeItem[n][m];

			\qquad \qquad \} \\

			\qquad \qquad found = stringCodeItem.find(stringTruncItem); \\

			\qquad \qquad if (found $\geq$ 0)

			\qquad \qquad \{

				\qquad \qquad \qquad uniqueCode[k] = uniqueCode[k] + 1;

				\qquad \qquad \qquad corrSymbolLocations[k] = n;

			\qquad \qquad \}

		\qquad \}

	\} \\

	//identifying the messageLevel value

	for (k = 1; k $\leq$ decodedMessageLength; k++)  

	\{

		\qquad if (uniqueCode[k] $>$ 1)

		\qquad \{

			\qquad \qquad messageLevel = 2;

		\qquad \}

	\} \\

	//identifying the messageLevel value

	for (k = 1; k $\leq$ decodedMessageLength; k++)  

	\{

		\qquad if (uniqueCode[k] == 0)

		\qquad \{

			\qquad \qquad messageLevel = 0;

		\qquad \}

	\} \\

	if (messageLevel == 1)

	\{

		\qquad for (k = 1; k $\leq$ decodedMessageLength; k++)  

		\qquad \{

			\qquad \qquad for (n = 1; n $\leq$ 40; n++)

			\qquad \qquad \{

				\qquad \qquad \qquad if (n == corrSymbolLocations[k])

				\qquad \qquad \qquad \{

					\qquad \qquad \qquad \qquad //forming the decoded message array

					\qquad \qquad \qquad \qquad charFinalDecodedMessage[k] = charAlphaNumeric[n]; 

				\qquad \qquad \qquad \}

			\qquad \qquad \} \\

			\qquad \qquad //forming the decoded message string

			\qquad \qquad stringFinalDecodedMessage=stringFinalDecodedMessage+charFinalDecodedMessage[k];

		\qquad \} \\

		\qquad outputFile $<$$<$ stringFinalDecodedMessage $<$$<$ endl $<$$<$ endl;

		\qquad outputFile $<$$<$ "The printed message has no known issues." $<$$<$ endl;

	\} \\

	if (messageLevel == 0)

	\{

		\qquad outputFile $<$$<$ "An incorrect code was used for decoding." $<$$<$ endl;

	\} \\

	if (messageLevel == 2)

	\{

		\qquad outputFile $<$$<$ "The initial encoded message needs to be resent." $<$$<$ endl;

	\}  \\

	codeFile.close();

	initialFile.close(); \\

	outputFile.close(); \\

	return 0;

\noindent
\} \\

\subsection{Breaking the Code Program} \label{Br}

\vspace{10pt} 

\noindent
\#include $<$iostream$>$

\noindent
\#include $<$fstream$>$

\noindent
\#include $<$string$>$

\noindent
\#include $<$stdio.h$>$

\noindent
\#include $<$stdlib.h$>$ // atof

\noindent
\#include $<$sstream$>$

\noindent
\#include $<$cmath$>$ \\

\noindent
using namespace std; \\

\noindent
int main() \{

	string secretCode;  //read from a file

	string initialCodedMessage;  //read from a file

	string stringCodeItem;  //temporary storage for a function value of an individual letter

	string stringTruncItem;  //temporary storage for a truncated function value of a coded message

	string stringRandomCodeItem;   //merged all truncated function values of a coded message

	string stringFinalDecodedMessage;  //final decoded message printed into output file

	const string alphanumeric = "?\_.,ABCDEFGHIJKLMNOPQRSTUVWXYZ1234567890"; \\

	char charAlphaNumeric[41];  //alphabet items

	char charFinalDecodedMessage[101];  //final decoded message items

	char charCodeItem[41][16];  //function value for an individual letter

	char charRandomCodeItem[101][10];  //truncated function value for an individual letter \\

	int codeLength;  //lenght of the secretCode

	int codedMessageLenght;  //lenght of the initialCodedMessage

	int codedShortMessageLenght;  //lenght of the stringRandomCodeItem

	int decodedMessageLength;  //lenght of the decoded message

	int codeSum;  //sum of the code digits 

	int firstFib;  //first number to start Fib

	int secondFib;  //second number to start Fib

	int extraSum;  //sum of the misleading items

	int found;   //tells at what position if any a substring was found in a string

	int messageLevel;  //tells if the message was decoded correctly \\

	int initialCode[11];  //array for initial numerical secret code

	int completedFullCode[201];  //array for completed full numerical secret code

	int completedOddFullCode[101];  //array for completed odd full numerical secret code

	int completedEvenFullCode[101];  //array for completed even full numerical secret code

	int corrSymbolLocations[101];  //array of correct locations of symbols in the final decoded message

	int uniqueCode[101];  //tells if the message was decoded in a unique way  \\

	long double numberCode;  //numerical secret code

	long double modifiedCode;  //numerical secret code with a shorter range

	long double sumItem;  //part of the major function

	long double diffValue;  //helps to find codeSum \\

	long long startPoint;   //first value of the code to be tried less than finishPoint

	long long finishPoint;    //last value of the code to be tried less than pow(10, codeLength)

	long long numberOfIterations;  //counts the number of iterations in the while loop \\

	long double codeItem[41];  //function value for an individual letter \\

	bool uniqueCodeIdentifyer; //tells if the message was decoded in a unique way \\

	ifstream initialFile; \\

	ofstream outputFile; \\

	initialFile.open("Initial\_Coded\_Message.txt"); \\

	outputFile.open("Final\_Decoded\_Message.txt"); \\

	getline(initialFile, initialCodedMessage); \\

	codeLength = 10;

	startPoint = 1234567800;

	finishPoint = startPoint + 100;

	numberOfIterations = 0.0; \\

	int k = 0;

	int j = 0;

	int n = 0;

	int m = 0; \\

	codedMessageLenght = initialCodedMessage.length(); \\

	for (k = 1; k $\leq$ 40; k++)  //forming function value for an individual letter

	\{

		\qquad charAlphaNumeric[k] = alphanumeric[k - 1];

	\} \\

	while (numberOfIterations $\leq$ (finishPoint - startPoint))  //all the iterations will be performed

	\{

		\qquad stringCodeItem = "";

		\qquad stringRandomCodeItem = "";

		\qquad stringTruncItem = "";

		\qquad stringFinalDecodedMessage = ""; \\

		\qquad codedShortMessageLenght = 0;

		\qquad decodedMessageLength = 0;

		\qquad codeSum = 0;

		\qquad firstFib = 0;

		\qquad secondFib = 0;

		\qquad extraSum = 0;

		\qquad found = -1;

		\qquad messageLevel = 1; \\

		\qquad numberCode = 0.0;

		\qquad modifiedCode = 0.0;

		\qquad sumItem = 0.0;

		\qquad diffValue = 0.0; \\

		\qquad uniqueCodeIdentifyer = true; \\

		\qquad for (k = 0; k $\leq$ 10; k++)

		\qquad \{

			\qquad \qquad initialCode[k] = 0;

		\qquad \} \\

		\qquad for (k = 0; k $\leq$ 40; k++)

		\qquad \{

			\qquad \qquad codeItem[k] = 0; \\

			\qquad \qquad for (j = 0; j $\leq$ 15; j++)

			\qquad \qquad \{

				\qquad \qquad \qquad charCodeItem[k][j] = ' ';

			\qquad \qquad \}

		\qquad \} \\

		\qquad for (k = 0; k $\leq$ 100; k++)

		\qquad \{

			\qquad \qquad completedOddFullCode[k] = 0;

			\qquad \qquad completedEvenFullCode[k] = 0;

			\qquad \qquad corrSymbolLocations[k] = 0;

			\qquad \qquad uniqueCode[k] = 0; \\

			\qquad \qquad for (j = 0; j $\leq$ 9; j++)

			\qquad \qquad \{

				\qquad \qquad \qquad charRandomCodeItem[k][j] = ' ';

			\qquad \qquad \} \\

			\qquad \qquad charFinalDecodedMessage[k] = ' ';

		\qquad \} \\

		\qquad for (k = 0; k $\leq$ 200; k++)

		\qquad \{

			\qquad \qquad completedFullCode[k] = 0;

		\qquad \} \\

		\qquad numberCode = startPoint + numberOfIterations; \\

		\qquad if (numberCode $<$ 1 * pow(10, codeLength - 1))

		\qquad \{

			\qquad \qquad modifiedCode = numberCode + pow(10, codeLength - 1);

		\qquad \}

		\qquad else

		\qquad \{

			\qquad \qquad if (numberCode $>$ 9 * pow(10, codeLength - 1))

			\qquad \qquad \{

				\qquad \qquad \qquad modifiedCode = numberCode - pow(10, codeLength - 1);

			\qquad \qquad \}

			\qquad \qquad else

			\qquad \qquad \{

				\qquad \qquad \qquad modifiedCode = numberCode;

			\qquad \qquad \}

		\qquad \} \\

		\qquad modifiedCode = modifiedCode / pow(10, codeLength) + 1; \\

		\qquad for (k = 1; k $\leq$ 40; k++)

		\qquad \{

			\qquad \qquad sumItem = sumItem + pow(modifiedCode, k);

		\qquad \} \\

		\qquad for (k = 1; k $\leq$ 40; k++)

		\qquad \{

			\qquad \qquad codeItem[k] = pow(modifiedCode, k) * pow(10, 30) / sumItem; 

			\qquad \qquad stringCodeItem = to\_string(codeItem[k]); \\

			\qquad \qquad for (j = 1; j $\leq$ 15; j++)

			\qquad \qquad \{

				\qquad \qquad \qquad charCodeItem[k][j] = stringCodeItem[j - 1];

			\qquad \qquad \}

		\qquad \} \\

		\qquad //forming completedOddFullCode and completedEvenFullCode

		\qquad for (k = 1; k $\leq$ codeLength; k++)  

		\qquad \{

			\qquad \qquad initialCode[k] = numberCode / pow(10, codeLength - k) - diffValue;

			\qquad \qquad diffValue = (diffValue + initialCode[k]) * 10;

			\qquad \qquad codeSum = codeSum + initialCode[k];

		\qquad \} \\

		\qquad firstFib = (codeSum \% 100) / 10;

		\qquad secondFib = codeSum \% 10; \\

		\qquad completedFullCode[1] = firstFib;

		\qquad completedFullCode[2] = secondFib; \\

		\qquad for (k = 3; k $\leq$ 200; k++)

		\qquad \{

			\qquad \qquad completedFullCode[k] = (completedFullCode[k - 1] + completedFullCode[k - 2]) \% 10;

		\qquad \} \\

		\qquad for (k = 1; k $\leq$ 200; k++)

		\qquad \{

			\qquad \qquad if (completedFullCode[k] $<$ 5)

			\qquad \qquad \{

				\qquad \qquad \qquad completedFullCode[k] = 9 - completedFullCode[k];

			\qquad \qquad \}

		\qquad \} \\

		\qquad for (k = 1; k $\leq$ 100; k++)

		\qquad \{

			\qquad \qquad completedOddFullCode[k] = completedFullCode[k * 2 - 1];

			\qquad \qquad completedEvenFullCode[k] = completedFullCode[k * 2];

		\qquad \} \\

		\qquad for (k = 1; k $\leq$ 100; k++) //forming a string with concatenated trancated real items

		\qquad \{

			\qquad \qquad for (n = 1; n $\leq$ completedOddFullCode[k]; n++)

			\qquad \qquad \{

				\qquad \qquad \qquad if (n + extraSum $\leq$ codedMessageLenght)

				\qquad \qquad \qquad \{

					\qquad \qquad \qquad \qquad stringRandomCodeItem=stringRandomCodeItem+initialCodedMessage[n+extraSum-1];

				\qquad \qquad \qquad \}

			\qquad \qquad \} \\

			\qquad \qquad extraSum = extraSum + completedOddFullCode[k] + completedEvenFullCode[k];

		\qquad \} \\

		\qquad codedShortMessageLenght = stringRandomCodeItem.length();

		\qquad extraSum = 0; \\

		\qquad for (k = 1; k $\leq$ 100; k++) //forming charRandomCodeItem[][]

		\qquad \{

			\qquad \qquad for (n = 1; n $\leq$ completedOddFullCode[k]; n++)

			\qquad \qquad \{

				\qquad \qquad \qquad if (n + extraSum $\leq$ codedShortMessageLenght)

				\qquad \qquad \qquad \{

					\qquad \qquad \qquad \qquad charRandomCodeItem[k][n] = stringRandomCodeItem[n + extraSum - 1];

				\qquad \qquad \qquad \}

			\qquad \qquad \} \\

			\qquad \qquad if (completedOddFullCode[k] + extraSum $\leq$ codedShortMessageLenght)

			\qquad \qquad \{

				\qquad \qquad \qquad decodedMessageLength = decodedMessageLength + 1;

			\qquad \qquad \} \\

			\qquad \qquad extraSum = extraSum + completedOddFullCode[k];

		\qquad \} \\

		\qquad for (k = 1; k $\leq$ decodedMessageLength; k++)  //finding the matching symbol

		\qquad \{

			\qquad \qquad stringTruncItem = ""; \\

			\qquad \qquad for (j = 1; j $\leq$ completedOddFullCode[k]; j++)

			\qquad \qquad \{

				\qquad \qquad \qquad stringTruncItem = stringTruncItem + charRandomCodeItem[k][j];

			\qquad \qquad \} \\

			\qquad \qquad for (n = 1; n $\leq$ 40; n++)

			\qquad \qquad \{

				\qquad \qquad \qquad stringCodeItem = ""; \\

				\qquad \qquad \qquad for (m = 1; m $\leq$ 15; m++)

				\qquad \qquad \qquad \{

					\qquad \qquad \qquad \qquad stringCodeItem = stringCodeItem + charCodeItem[n][m];

				\qquad \qquad \qquad \} \\

				\qquad \qquad \qquad found = stringCodeItem.find(stringTruncItem); \\

				\qquad \qquad \qquad if (found $\geq$ 0)

				\qquad \qquad \qquad \{

					\qquad \qquad \qquad \qquad uniqueCode[k] = uniqueCode[k] + 1;

					\qquad \qquad \qquad \qquad corrSymbolLocations[k] = n;

				\qquad \qquad \qquad \} \\

			\qquad \qquad \}

		\qquad \} \\

		\qquad  //identifying the messageLevel value

		\qquad for (k = 1; k $\leq$ decodedMessageLength; k++) 

		\qquad \{

			\qquad \qquad if (uniqueCode[k] $>$ 1)

			\qquad \qquad \{

				\qquad \qquad \qquad messageLevel = 2;

			\qquad \qquad \}

		\qquad \} \\

		\qquad //identifying the uniqueCodeIdentifyer and messageLevel value

		\qquad for (k = 1; k $\leq$ decodedMessageLength; k++) 

		\qquad \{

			\qquad \qquad if (uniqueCode[k] == 0)

			\qquad \qquad \{

				\qquad \qquad \qquad uniqueCodeIdentifyer = false;

				\qquad \qquad \qquad messageLevel = 0;

			\qquad \qquad \}

		\qquad \} \\

		\qquad if (uniqueCodeIdentifyer == true)

		\qquad \{

			\qquad \qquad //forming the decoded message array

			\qquad \qquad for (k = 1; k $\leq$ decodedMessageLength; k++) 

			\qquad \qquad \{

				\qquad \qquad \qquad for (n = 1; n $\leq$ 40; n++)

				\qquad \qquad \qquad \{

					\qquad \qquad \qquad \qquad if (n == corrSymbolLocations[k])

					\qquad \qquad \qquad \qquad \{

						\qquad \qquad \qquad \qquad \qquad charFinalDecodedMessage[k] = charAlphaNumeric[n];

					\qquad \qquad \qquad \qquad \}

				\qquad \qquad \qquad \}

			\qquad \qquad \} \\

			\qquad \qquad //forming the decoded message string

			\qquad \qquad for (k = 1; k $\leq$ decodedMessageLength; k++) 

			\qquad \qquad \{

				\qquad \qquad \qquad stringFinalDecodedMessage=stringFinalDecodedMessage+charFinalDecodedMessage[k];

			\qquad \qquad \} \\

			\qquad \qquad outputFile $<$$<$ stringFinalDecodedMessage $<$$<$ endl $<$$<$ endl; \\

			\qquad \qquad secretCode = to\_string(startPoint + numberOfIterations); \\

			\qquad \qquad if (messageLevel == 1)

			\qquad \qquad \{

				\qquad \qquad \qquad outputFile $<$$<$ "The secret code was determined successfully." $<$$<$ endl;

				\qquad \qquad \qquad outputFile $<$$<$ "The printed message has no known issues." $<$$<$ endl;

				\qquad \qquad \qquad outputFile $<$$<$ "The secret code is: " $<$$<$ secretCode $<$$<$ endl;

			\qquad \qquad \} \\

			\qquad \qquad if (messageLevel == 2)

			\qquad \qquad \{

				\qquad \qquad \qquad outputFile $<$$<$ "The secret code was determined." $<$$<$ endl;

				\qquad \qquad \qquad outputFile $<$$<$ "However, the printed message might have some issues." $<$$<$ endl;

				\qquad \qquad \qquad outputFile $<$$<$ "The hypothetical secret code is: " $<$$<$ secretCode $<$$<$ endl;

			\qquad \qquad \} \\

			\qquad \qquad numberOfIterations = finishPoint - startPoint; 

		\qquad \} \\

		\qquad numberOfIterations = numberOfIterations + 1.0;

	\} \\

	if (uniqueCodeIdentifyer == false)

	\{

		\qquad outputFile $<$$<$ "The secret code was not determined." $<$$<$ endl;

		\qquad outputFile $<$$<$ "Try a different range." $<$$<$ endl;

	\} \\

	initialFile.close(); \\

	outputFile.close(); \\

	return 0;

\noindent
\} \\


\end{document}